\documentclass[12pt,aps,prb,preprint]{revtex4}
\usepackage{amsmath}
\usepackage{graphicx}

 \def\be{\begin{equation}}
 \def\ee{\end{equation}}
 \def\bea{\begin{eqnarray}}
 \def\eea{\end{eqnarray}}
 \def\bc{\begin{center}}
 \def\ec{\end{center}}

 \def\etal{ et al.}
 \newcommand{\ind}{\hspace*{\parindent}}
 \newcommand{\vs}{\vskip .1in}
 
 \newcommand{\doublespace}{
         \renewcommand{\baselinestretch}{1.75}
         \large\normalsize}
\usepackage{epsfig}
\usepackage{graphicx} % Include figure files
\usepackage{dcolumn}  % Align table columns on decimal point
\usepackage{bm}       % bold math
\usepackage{amsfonts, amssymb}

\doublespace
\begin{document}

\title{Paradoxical pop-ups:  Why are they hard to catch?}

\author{Michael K. McBeath}
\email{m.m@asu.edu} \affiliation{Department of Psychology, Arizona
State University, Tempe, AZ 85287}

\author{Alan M. Nathan}
\email{a-nathan@uiuc.edu} \affiliation{Department of Physics,
University of Illinois, Urbana, Illinois 61801}

\author{A. Terry Bahill}
\email{terry@sie.arizona.edu} \affiliation{Department of Systems
and Industrial Engineering, University of Arizona, Tucson, AZ
85721}

\author{David G. Baldwin}
\email{snakejazz_37@brainpip.com} \affiliation{PO Box 190,
Yachats, OR 97498}

\begin{abstract}

Even professional baseball players occasionally find it difficult
to gracefully approach seemingly routine pop-ups.  This paper
describes a set of towering pop-ups with trajectories that exhibit
cusps and loops near the apex.  For a normal fly ball, the
horizontal velocity is continuously decreasing due to drag caused
by air resistance. But for pop-ups, the Magnus force (the force
due to the ball spinning in a moving airflow) is larger than the
drag force.  In these cases the horizontal velocity decreases in
the beginning, like a normal fly ball, but after the apex, the
Magnus force accelerates the horizontal motion. We refer to this
class of pop-ups as paradoxical because they appear to misinform
the typically robust optical control strategies used by fielders
and lead to systematic vacillation in running paths, especially
when a trajectory terminates near the fielder.  In short, some of
the dancing around when infielders pursue pop-ups can be well
explained as a combination of bizarre trajectories and misguidance
by the normally reliable optical control strategy, rather than
apparent fielder error.  Former major league infielders confirm
that our model agrees with their experiences.

\end{abstract}

\maketitle

\section{ Introduction}
\label{sec:intro}

Baseball has a rich tradition of misjudged pop-ups.  For example,
in April, 1961, Roy Sievers of the Chicago White Sox hit a
towering pop-up above Kansas City Athletics' third baseman Andy
Carey who fell backward in trying to make the catch.  The ball
landed several feet from third base, far out of the reach of
Carey.  It rolled into the outfield, and Sievers wound up on
second with a double.

A few other well-known misplays of pop-ups
include: New York Giants' first baseman Fred Merkle's failure to
catch a foul pop-up in the final game of the 1912 World Series,
costing the Giants the series against the Boston Red Sox; St.
Louis first baseman Jack Clark's botched foul pop-up in the sixth
game of the 1985 World Series against Kansas City; and White Sox
third baseman Bill Melton's broken nose suffered in an attempt to
catch a ``routine'' pop-up in 1970.

As seen by these examples,
even experienced major league baseball players can find it
difficult to position themselves to catch pop-ups hit very high
over the infield.  Players describe these batted balls as
``tricky'' or ``deceptive,'' and at times they will be seen
lunging for the ball in the last instant of the ball's descent.
``Pop-ups look easy to anyone who hasn't tried to catch one - like
a routine fly ball that you don't have to run for,'' Clete Boyer
said, ``but they are difficult to judge and can really make you
look like an idiot.'' Boyer, a veteran of sixteen years in the
major leagues, was considered one of the best defensive infielders
in baseball.

Several factors can exacerbate the infielder's
problem of positioning himself for a pop-up.  Wind currents high
above the infield can change the trajectory of the pop-up
radically.  Also, during day games the sky might provide little
contrast as a background for the ball--a condition called a ``high
sky'' by players.  Then, there are obstacles on the field--bases,
the pitcher's mound, and teammates--that can hinder the infielder
trying to make a catch.  But even on a calm night with no
obstacles nearby, players might stagger in their efforts to get to
the ball.

The frequency of pop-ups in the major leagues--an
average of nearly five pop-ups per game--is great enough that
teams provide considerable pop-up practice for infielders and
catchers. Yet, this practice appears to be severely limited in
increasing the skill of these players.  Infielders seem unable to
reach the level of competency in catching ``sky-high'' pop-ups
that outfielders attain in catching high fly balls, for example.
This suggests that the technique commonly used to catch pop-ups
might be the factor limiting improvement.

Almost all baseball players learn to catch low, ``humpback''
pop-ups and fly balls before they have any experience in catching
lofty pop-ups.  In youth leagues nearly all pop-ups have low
velocities and few exceed a height of fifty feet; therefore, they
have trajectories that are nearly parabolic.  Fly balls, too, have
near-parabolic trajectories. Young players develop techniques for
tracking low pop-ups and fly balls.  If 120-foot pop-ups do not
follow similar trajectories, however, major league infielders
might find pop-ups are hard to catch because the tracking and
navigation method they have learned in their early years is
unreliable for high, major league pop-ups.

In the consideration of this hypothesis, we first describe
trajectories of a set of prototypical batted balls, using models
of the bat-ball collision and ball flight aerodynamics.  We then
develop models of three specific kinds of typical non-parabolic
pop-up trajectories.  These ``paradoxical'' trajectories exhibit
unexpected behavior around their apices, including cusps and
loops.  Several of these paradoxical trajectories are fitted with
an optical control model that has been used successfully to
describe how players track and navigate to fly balls. For each
fit, a prediction of the behavior of infielders attempting to
position themselves to catch high pop-ups is compared with the
observed behavior of players during games.

\section{Simulations of batted-ball trajectories}
\label{sec:simulations}
\subsection{Forces on a spinning baseball in flight}
\begin{figure}[t]
\epsfig{width=2in,file=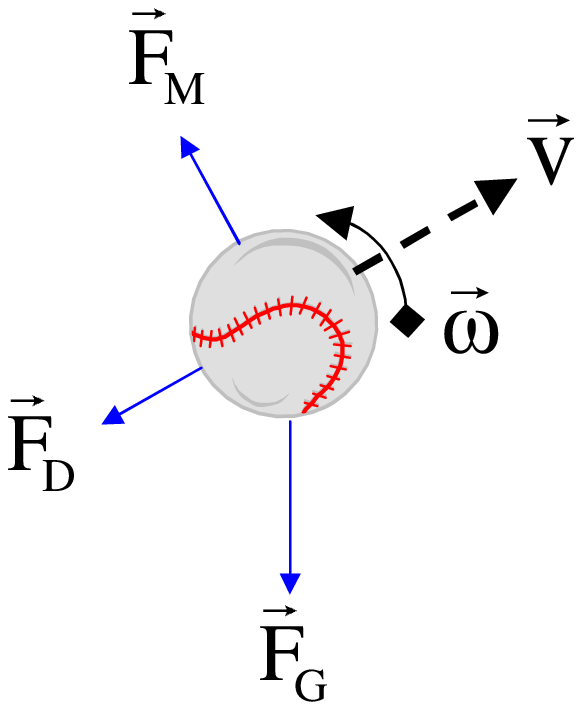} \caption {\small  The
forces on a baseball in flight with backspin, including gravity
($F_G$), drag ($F_D$), and the Magnus force ($F_M$). $F_D$ acts in
the $-\hat{v}$ direction and $F_M$ acts in the
$\hat{\omega}\times\hat{v}$ direction.} \label{fig:forces}
\end{figure}

\ind As every student in an introductory physics course learns,
the trajectory of a fly ball in a vacuum is a smooth symmetric
parabola since the only force acting on it is the downward pull of
gravity.   However, in the atmosphere the ball is subject to
additional forces, shown schematically in Fig.~\ref{fig:forces}:
the retarding force of drag ($F_D$) and the Magnus force ($F_M$).
The Magnus force was first mentioned in the scientific literature
by none other than a young Isaac Newton in his treatise on the
theory of light,\cite{newton1671} where he included a brief
description on the curved trajectory of a spinning tennis ball.
Whereas the drag force always acts opposite to the instantaneous
direction of motion, the Magnus force is normal to both the
velocity and spin vectors. For a typical fly ball to the outfield,
the drag force causes the trajectory to be somewhat asymmetric,
with the falling angle steeper than the rising
angle,\cite{adair02} although the trajectory is still smooth.  If
the ball has backspin, as expected for such fly balls, the Magnus
force is primarily in the upward direction, resulting in a
higher--but still quite smooth--trajectory. However, as we will
show the situation is qualitatively very different for a pop-up,
since a ball-bat collision resulting in a pop-up will have a
considerable backspin, resulting in a significantly larger Magnus
force than for a fly ball.  Moreover, the direction of the force
is primarily horizontal with a sign that is opposite on the upward
and downward paths.   These conditions will result in unusual
trajectories--sometimes with cusps, sometimes with loops--that we
label as ``paradoxical.''

With this brief introduction, we next discuss our simulations of
baseball trajectories in which a model for the ball-bat collision
(Sec.~\ref{sec:collision}) is combined with a model for the drag
and Magnus forces (Sec.~\ref{sec:aero}) to produce the batted-ball
trajectories.  We discuss the paradoxical nature of these
trajectories in Sec~\ref{sec:trajs} in light of the interplay
among the various forces acting on the ball.

\subsection{Ball-bat collision model}
\label{sec:collision}

\begin{figure}[t]
\epsfig{width=3in,file=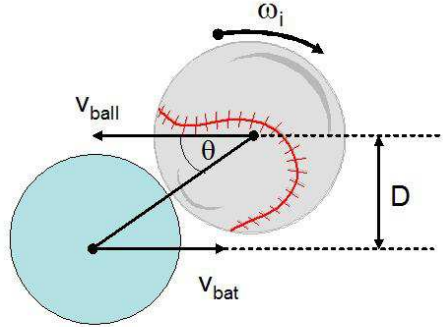} \caption {\small Geometry
of the ball-bat collision.  The initial velocity of the ball and
bat are $v_{ball}$ and $v_{bat}$, respectively, and the pitched
ball has backspin of magnitude $\omega_i$.  The bat-ball offset
shown in the figure is denoted by
$D=(r_{ball}+r_{bat})\sin\theta$, where $r_{ball}$ and $r_{bat}$
are the radii of the ball and bat, respectively.   For the
collisions discussed in the text, the entire picture should be
rotated counterclockwise by 8.6$^\circ$, so that the initial angle
of the ball is 8.6$^\circ$ downward and of the bat is 8.6$^\circ$
upward.} \label{fig:geom}
\end{figure}

\begin{figure}[t]
\epsfig{width=2in,angle=270,file=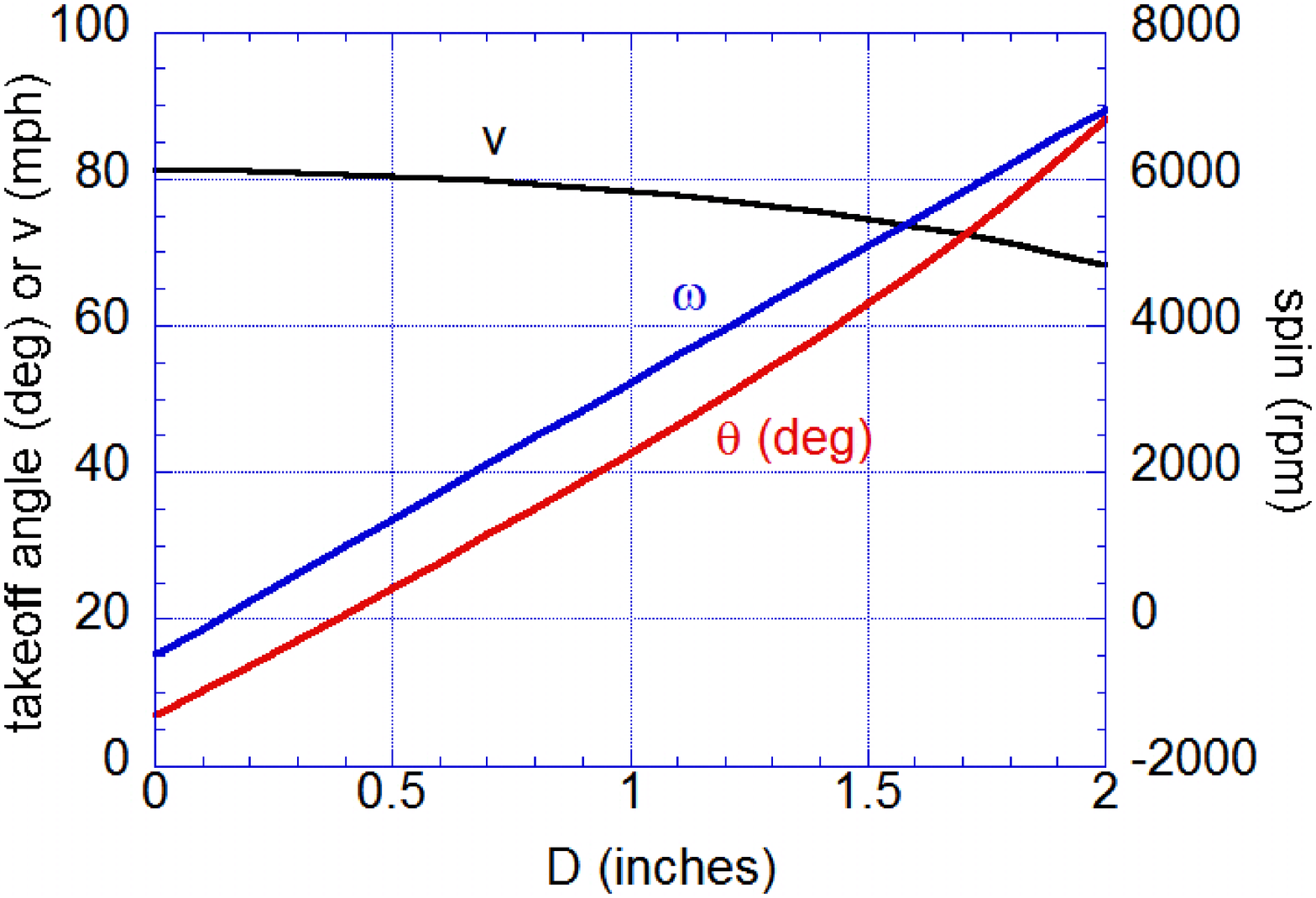} \caption{\small
Variation of the batted ball speed, initial angle above the
horizontal, and spin with the offset $D$.} \label{fig:d}
\end{figure}

The collision model is identical to that used both by
Sawicki\etal\cite{sawicki03} and by Cross and
Nathan.\cite{cross06} The geometry of the collision is shown in
Fig.~\ref{fig:geom}.  A standard baseball ($r_{ball}$=1.43 inch,
mass=5.1 oz) approaches the bat with an initial speed
$v_{ball}$=85 mph, initial backspin $\omega_i$=126 rad/s (1200
rpm), and at a downward angle of 8.6$^\circ$ (not shown in the
figure). The bat has an initial velocity $v_{bat}$=55 mph at the
point of impact and an initial upward angle of 8.6$^\circ$,
identical to the downward angle of the ball. The bat was a 34-inch
long, 32-oz wood bat with an R161 profile, with radius
$r_{bat}$=1.26 inch at the impact point.  If lines passing through
the center of the ball and bat are drawn parallel to the initial
velocity vectors, then those lines are offset by the distance $D$.
Simply stated, $D$ is the amount by which the bat undercuts
($D>0$) or overcuts ($D<0$) the ball.   In the absence of initial
spin on the baseball, a head-on collision ($D=0$) results in the
ball leaving the bat at an upward angle of 8.6$^\circ$ and with no
spin; undercutting the ball produces backspin and a larger upward
angle; overcutting the ball produces topspin and a smaller upward
or even a downward angle. The ball-bat collision is characterized
by two constants, the normal and tangential coefficients of
restitution--$e_N$ and $e_T$, respectively--with the additional
assumption that angular momentum is conserved about the initial
contact point between the ball and bat.\cite{cross06} For $e_N$,
we use the parameterization \be
 e_N  =  0.54 - (v_N-60)/895 \, ,
 \ee
where $v_N=(v_{ball}+v_{bat})\cos\theta$  is the normal component
of the relative ball-bat velocity in units of mph.\cite{sawicki03}
We further assume $e_T=0$, which is equivalent to assuming that
the tangential component of the relative ball-bat surface
velocity, initially equal to
$(v_{ball}+v_{bat})\sin\theta+r_{ball}\omega_i$, is identically
zero as the ball leaves the bat, implying that the ball leaves the
bat in a rolling motion.   The loss of tangential velocity occurs
as a result of sliding friction, and it was verified by direct
calculation that the assumed coefficient of friction of
0.55\cite{cross06} is sufficient to bring the tangential motion to
a halt prior to the end of the collision for all values of $D<1.7$
inches. Given the initial velocities and our assumptions about
$e_N$ and $e_T$, the outgoing velocity $v$, angle $\theta$ , and
backspin of the baseball can be calculated as a function of the
offset $D$. These parameters, which are shown in Fig.~\ref{fig:d},
along with the initial height of 3 ft., serve as input into the
calculation of the batted-ball trajectory. Note particularly that
both $\omega$  and $\theta$ are strong functions of $D$, whereas
$v$ only weakly depends on $D$.

\subsection{Baseball aerodynamics model}
\label{sec:aero}
 The trajectory of the batted baseball is
calculated by numerically solving the differential equations of
motion using a fourth-order Runge-Kutta technique, given the
initial conditions and the forces.  Conventionally, drag and
Magnus forces are written as \bea
\vec{F_D}&=&-\frac{1}{2}C_D\rho Av^2\, \hat{v}\\
\vec{F_M}&=&\frac{1}{2}C_L\rho Av^2\, (\hat{\omega}\times\hat{v})
\, , \label{eq:drag} \eea where $\rho$ is the air density (0.077
lb/ft$^3$), $A$ is the cross sectional area of the ball (6.45
inch$^2$), $v$ is the velocity, $\omega$ is the angular velocity,
and $ C_D$ and $C_L$ are phenomenological drag and lift
coefficients, respectively.   Note that the direction of the drag
is opposite to the direction of motion whereas the direction of
the Magnus force is determined by a right-hand rule. We utilize
the parametrizations of Sawicki et al.\cite{sawicki03} in which
$C_D$ is a function of the speed $v$ and $C_L$ is a bilinear
function of spin parameter $S=r_{ball}/v$, implying that $F_M$ is
proportional to $\omega v$.   Since the velocity of the ball does
not remain constant during the trajectory, it is necessary to
recompute $C_D$ and $C_L$ at each point in the numerical
integration.
%However, our calculations assumed constant
%$\omega$, in accord with the analysis of Sawicki et
%al.\cite{sawicki05}
The resulting trajectories are shown in
Fig.~\ref{fig:trajs} for values of $D$ in the range 0-1.7 inches,
where an initial height of 3 ft was assumed.

\subsection{Discussion of trajectories}
\label{sec:trajs}

\begin{figure}[t]
\epsfig{width=3in,file=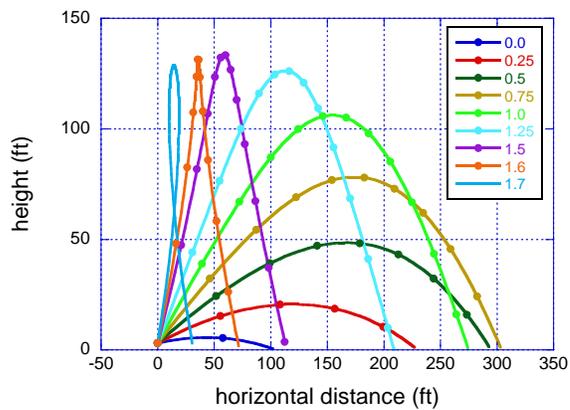} \caption{\small Simulated
trajectories of pop-ups, fly balls, and line drives with drag and
spin-induced forces. These trajectories were produced when an 85
mph fastball with 1200 revolutions per minute (rpm) backspin
collided with the sweet spot of a bat that was moving at 55 mph.
Each trajectory was created by a different offset $D $(in inches)
between the bat and ball, as defined in Fig.~\ref{fig:geom}.}
\label{fig:trajs}
\end{figure}

The striking feature of Fig.~\ref{fig:trajs}  is the qualitatively
different character of the trajectories as a function of $D$, or
equivalently as a function of the takeoff angle $\theta$.   These
trajectories range from line drives at small $\theta$, to fly
balls at intermediate $\theta$, to pop-ups at large $\theta$.
Particularly noteworthy is the rich and complex behavior of the
pop-ups, including cusps and loops.  The goal of this section is
to understand these trajectories in the context of the interplay
among the forces acting on the ball.  To our knowledge, there has
been no previous discussion of such unusual trajectories in the
literature.  We focus on two particular characteristics that may
have implications for the algorithm used by a fielder to catch the
ball:  the symmetry/asymmetry about the apex and the curvature.
Before proceeding, however, we remark that the general features of
the trajectories shown in Fig.~\ref{fig:trajs}  are universal and
do not depend on the particular model used for either the ball-bat
collision or for the drag and lift.  For example, using collision
and aerodynamics models significantly different from those used
here, Adair finds similar trajectories with both cusp-like and
loop-like behavior,\cite{adair02} which we verify with our own
calculations using his model.  Models based on equations in Watts
and Bahill\cite{watts00} result in similar trajectories.

We first examine the symmetry, or lack thereof, of the trajectory
about the apex.  Without the drag and Magnus forces, all
trajectories would be symmetric parabolas; the actual situation is
more complicated.  As seen in Fig.~\ref{fig:trajs}, baseballs hit
at low and intermediate  $\theta$ (line drives and fly balls) have
an asymmetric trajectory, with the ball covering less horizontal
distance on the way down than it did on the way up.  This feature
is known intuitively to experienced outfielders.  For larger
$\theta$ the asymmetry is smaller, and pop-ups hit at a very steep
angle are nearly symmetric.  How do the forces conspire to produce
these results?

\begin{figure}[t]
\epsfig{width=3in,angle=0,file=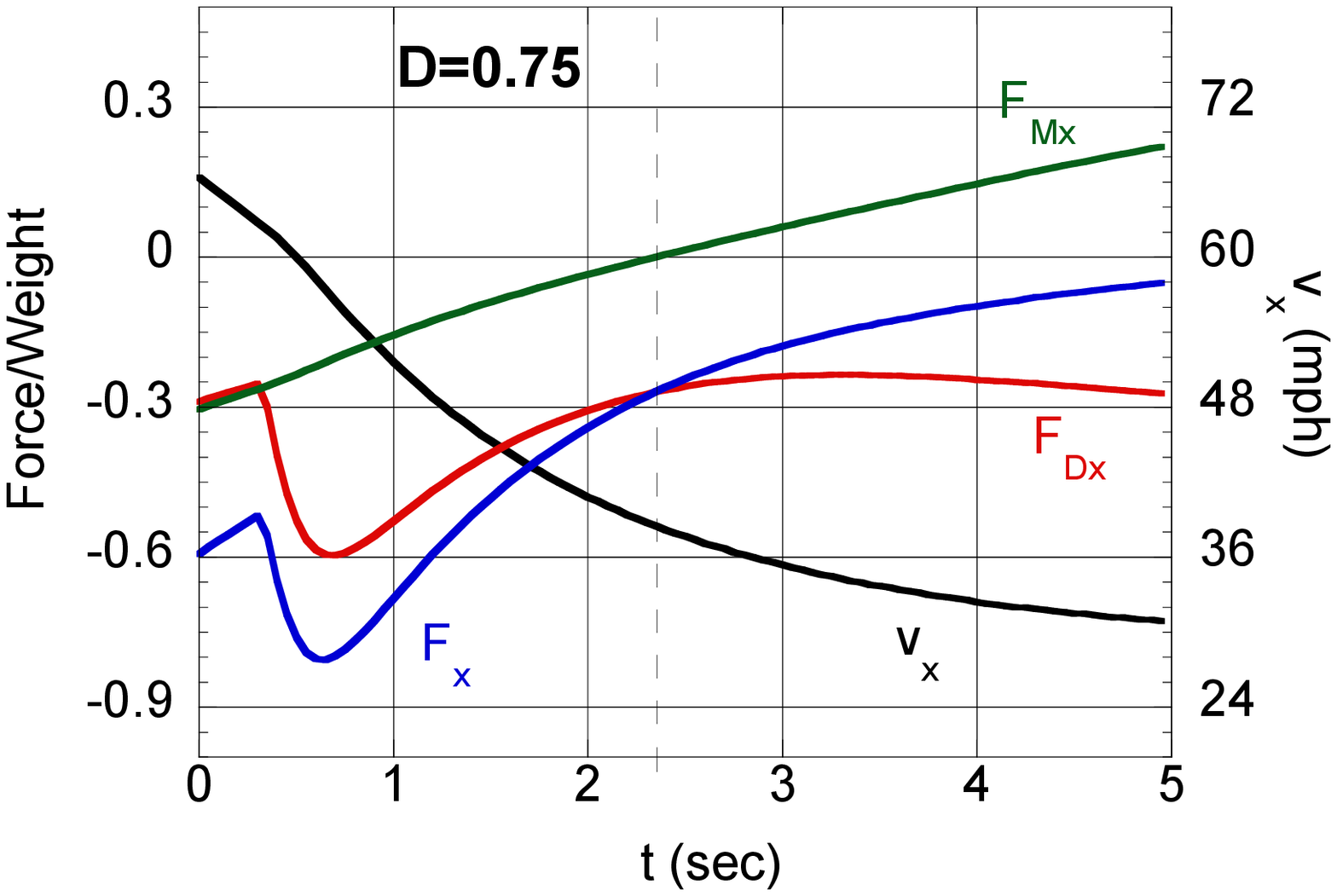} \caption{\small
Time dependence of the horizontal velocity, $v_x$, and the
horizontal forces for the $D=0.75$ trajectory, typical of a long
fly ball. The drag, lift, and total forces are denoted by
$F_{Dx}$, $F_{Mx}$, and $F_x$, respectively, normalized to the
weight.  The vertical dashed line indicates the time that the apex
is reached.} \label{fig:forcet1}
\end{figure}

\begin{figure}[ht]
\epsfig{width=3in,angle=0,file=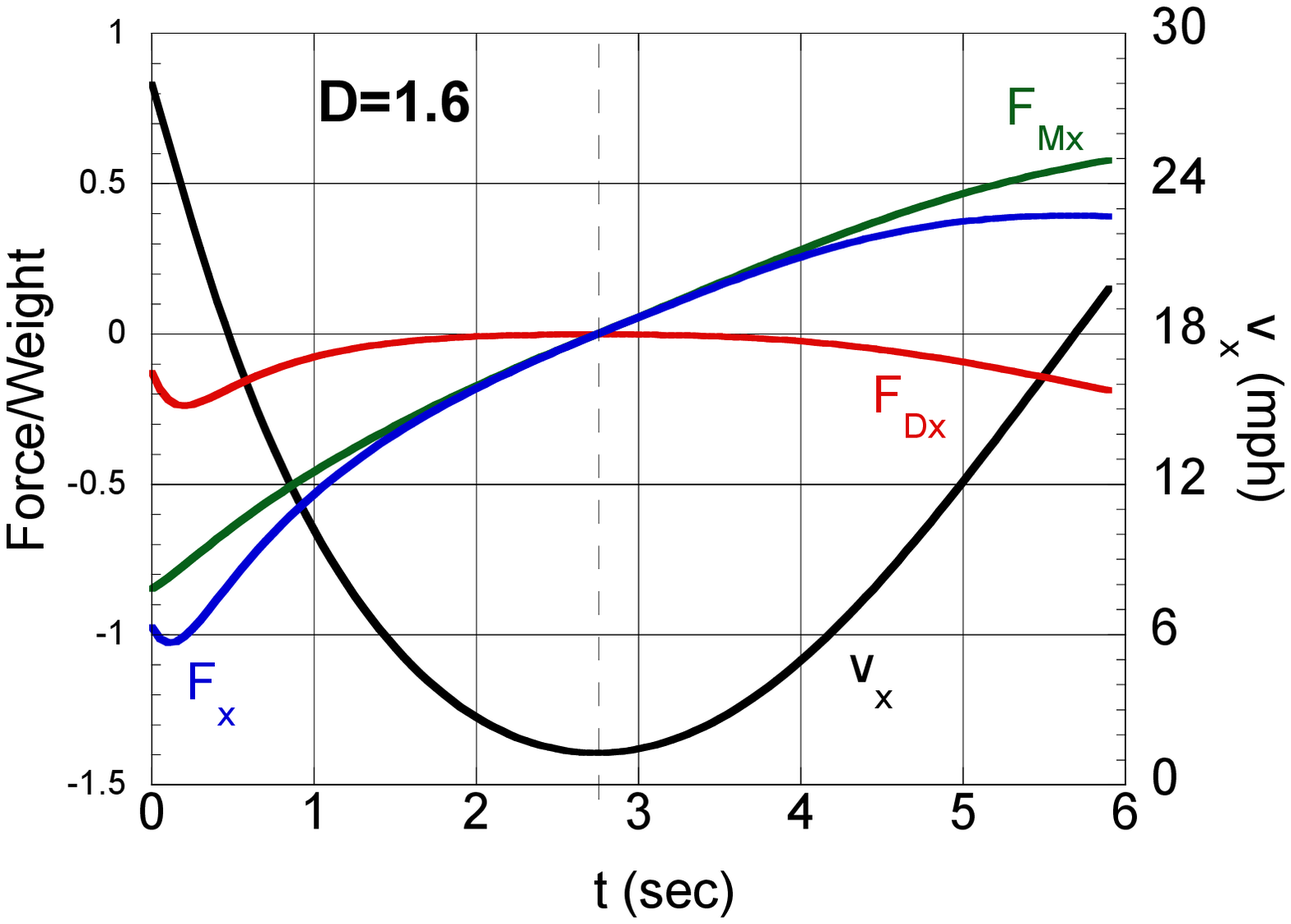} \caption{\small
Time dependence of the horizontal velocity, $v_x$, and the
horizontal forces for the $D=1.6$ trajectory, typical of a high
popup. The drag, lift, and total forces are denoted by $F_{Dx}$,
$F_{Mx}$, and $F_x$, respectively, normalized to the weight.  The
vertical dashed line indicates the time that the apex is reached.}
\label{fig:forcet2}
\end{figure}

We address this question by referring to Figs.~\ref{fig:forcet1}
and \ref{fig:forcet2}, in which the time dependence of the
horizontal components of the velocity and the forces are plotted
for a fly ball ($D=0.75$, $\theta=33^\circ$) and a pop-up
($D=1.6$, $\theta=68^\circ$). The initial decrease of the drag
force for early times is due to the particular model used for the
drag coefficient, which experiences a sharp drop near 75 mph.  The
asymmetry of the trajectory depends on the interplay between the
horizontal components of drag and Magnus, $F_{Dx}$ and $F_{Mx}$,
respectively.   For forward-going trajectories ($v_x>0$), $F_{Dx}$
always acts in the -x direction, whereas $F_{Mx}$ acts in the -x
or +x direction on the rising or falling part of the trajectory,
respectively.   The relative magnitudes of $F_{Dx}$ and $F_{Mx}$
depend strongly on both $\theta$ and $\omega$. For fly balls,
$\theta$  and  $\omega$ are small enough (see Fig.~\ref{fig:d})
that the magnitude of $F_{Dx}$ is generally larger than the
magnitude of $F_{Mx}$, as shown in Fig.~\ref{fig:forcet1}.
Therefore $F_x$ is negative throughout the trajectory. Under such
conditions, there is a smooth continuous decrease in $v_x$,
leading to an asymmetric trajectory, since the horizontal distance
covered prior to the apex is greater than that covered after the
apex.  The situation is qualitatively and quantitatively different
for pop-ups, since both  $\theta$ and $\omega$  are significantly
larger than for a fly ball.  As a result, the magnitude of
$F_{Mx}$ is much greater than the magnitude of $F_{Dx}$.  Indeed,
Fig.~\ref{fig:forcet2} shows that $F_x\approx F_{Mx}$, so that
$F_x$ acts in the -x direction before the apex and in the +x
direction after the apex. Therefore, the loss of $v_x$ while
rising is largely compensated by a gain in $v_x$ while falling,
resulting in near symmetry about the apex. Moreover, for this
particular trajectory the impulse provided by $F_x$ while rising
is nearly sufficient to bring $v_x$ to zero at the apex, resulting
in the cusp-like behavior.  For even larger values of $\theta$,
$F_x$ is so large that $v_x$ changes sign prior to the apex, then
reverses sign again on the way down, resulting in the
loop-the-loop pattern.

We next address the curvature of the trajectory, $C\equiv
d^2y/dx^2$, which is determined principally by the interplay
between the Magnus force $F_M$ and the component of gravity normal
to the trajectory $F_{GN}=F_G\cos\theta$. It is straightforward to
show that $C$ is directly proportional to the instantaneous value
of $(F_M-F_{GN})/(v_x^2\cos\theta)$ and in particular that the
sign of $C$ is identical to the sign of $F_M-F_{GN}$.   In the
absence of a Magnus force, the curvature is always negative, even
if drag is present.  An excellent example is provided by the
inverted parabolic trajectories expected in the absence of
aerodynamic forces.  The trajectories shown in
Fig.~\ref{fig:trajs} fall into distinct categories, depending on
the initial angle $\theta$.  For small enough $\theta$, $C$ is
negative throughout the trajectory.  Indeed, if C is initially
negative, then it is always negative, since $F_M$ is never larger
and $F_{GN}$ is never smaller than it is at t=0. For our
particular collision and aerodynamic model, the initial curvature
is negative for $\theta$ less than about $45^\circ$. For
intermediate $\theta$, $C$ is positive at the start and end of the
trajectory but experiences two sign changes, one before and one
after the apex.  The separation between the two sign changes
decreases as $\theta$ increases, until the two values coalesce at
the apex, producing a cusp.  For larger values of  $\theta$, $C$
is positive throughout the trajectory, resulting in loop-like
behavior such as the $D=1.7$ trajectory, where the sign of $v_x$
is initially positive, then changes to negative before the apex,
and finally changes to back positive after the apex.

\subsection{Does the spin remain constant?}
\label{sec:spin}
 All the simulations reported thus far assume that
the spin remains constant throughout the trajectory. Since the
spin plays such a major role in determining the character of the
trajectory, it is essential to examine the validity of that
assumption.   To our knowledge, there have been no experimental
studies on the spin decay of baseballs, but there have been two
such studies for golf, one by Smits and Smith\cite{smits} and one
by Tavares et al.\cite{tavares}  Tavares et al. propose a
theoretical model for the spin decay of a golf ball in which the
torque responsible for the decay is expressed as $R\rho AC_Mv^2$,
where $R$ is the radius of the ball and $C_M$ is the ``coefficient
of moment" which is given by $C_M = \beta R\omega/v$.  By equating
the torque to $Id\omega/dt$, where $I=0.4MR^2$ is the moment of
inertia, the spin decay constant $\tau$ can be expressed as \be
\tau \, \equiv \, \frac{1}{\omega}\frac{d\omega}{dt}\, = \, \left
[\frac{M}{R^2}\right] \frac{0.4}{\pi\rho\beta v} \, .
\label{eq:tau}\ee Using their measurements of $\tau$, Tavares et
al. determine $\beta\approx 0.012$, corresponding to  $\tau=20$
sec for $v$=100 mph. The measurements of Smits and Smith can be
similarly interpreted with  $\beta$=0.009, corresponding to
$\tau=25$ sec at 100 mph. To estimate the spin decay constant for
a baseball, we assume Eq.~\ref{eq:tau} applies, with $M/R^2$
scaled appropriately for a baseball and with all other factors the
same. Using $M/R^2$ = 2.31 and 2.49 oz/inch$^2$ for a golf ball
and baseball, respectively, the decay time for a baseball is about
8\% longer than for a golf ball, or 22-27 sec at 100 mph and
longer for smaller $v$.   A similar time constant for baseball was
estimated by Sawicki et al.,\cite{sawicki05} quite possibly using
the same arguments as we use here.  Since the trajectories
examined herein are in the air 7 sec or less, we conclude that our
results are not affected by the spin decay. Adair has suggested a
much smaller decay time, of order 5 sec,\cite{adair02} which does
not seem to be based on any experimental data.  A direct check of
our calculations shows that the qualitative effects depicted in
Fig.~\ref{fig:trajs} persist even with a decay time as short as 5
sec.

\section{Optical control model for tracking and navigating baseballs}
\subsection{Overview}
\label{sec:models}
 In a seminal article Seville Chapman\cite{chapman68} proposed an optical control model
for catching fly balls, today known as Optical Acceleration
Cancellation (OAC). Chapman examined the geometry of catching from
the perspective of a moving fielder observing an approaching
ballistic target that is traveling along a parabola. He showed
that in this case, the fielder can be guided to the destination
simply by selecting a running path that keeps the image of the
ball rising at a constant rate in a vertical image plane.
Mathematically, the tangent of the vertical optical angle to the
ball increases at a constant rate.  When balls are headed to the
side, other optical control strategies become
available.\cite{mcbeath95,sugar06b} However, in the current paper
we examine cases of balls hit directly toward the fielder, so we
will emphasize predictions of the OAC control mechanism.

Chapman assumed parabolic trajectories because of his (incorrect)
belief that the drag and Magnus forces have a negligible effect on
the trajectory.  Of course we now know that the effects of these
forces can be considerable, as discussed in Sec.~\ref{sec:trajs}.
Yet despite this initial oversight, numerous perception-action
catching studies confirm that fielders actually do appear to
utilize Chapman's type of optical control mechanism to guide them
to interception, and in particular OAC is the only mechanism that
has been supported for balls headed in the sagittal plane directly
toward fielders.\cite{babler93,mcbeath95,mcleod96,sugar06a}
Further support for OAC has been found with dogs catching Frisbees
as well as functioning mobile robots.\cite{shafer04,sugar06b}

Extensive research on the navigational behavior of baseball
players supports that perceptual judgment mechanisms used during
fly ball catching can generally be divided into two
phases.\cite{mcbeath95,shaffer05}  During the first phase, while
the ball is still relatively distant, ball location information is
largely limited to the optical trajectory (i.e. the observed
trajectory path of the image of the ball).  During the second or
final phase, other cues such as the increase in optical size of
the ball, and the stereo angle between the two eyes also become
available and provide additional information for final corrections
in fielder positioning and timing.  The control parameters in
models like OAC are optical angles from the fielder's perspective,
which help direct fielder position relative to the ongoing ball
position. Considerable work exploring and examining the final
phase of catching has been done by perception
scientists\cite{mazyn07,salve93} and some recent speculation has
been done by physicists.\cite{adair07}  Researchers generally
agree that the majority of fielder movement while catching balls
takes place during the first phase in which fielders approach the
destination region where the ball is headed. In the current work,
we focus on control models like OAC that guide fielder position
during the initial phase of catching.  Thus for example, we would
consider the famous play in which Jose Canseco allowed a ball to
bounce off of his head for a home run to be a catch, in that he
was guided to the correct location to intercept the ball.

\begin{figure}[ht]
\epsfig{width=5in,angle=0,file=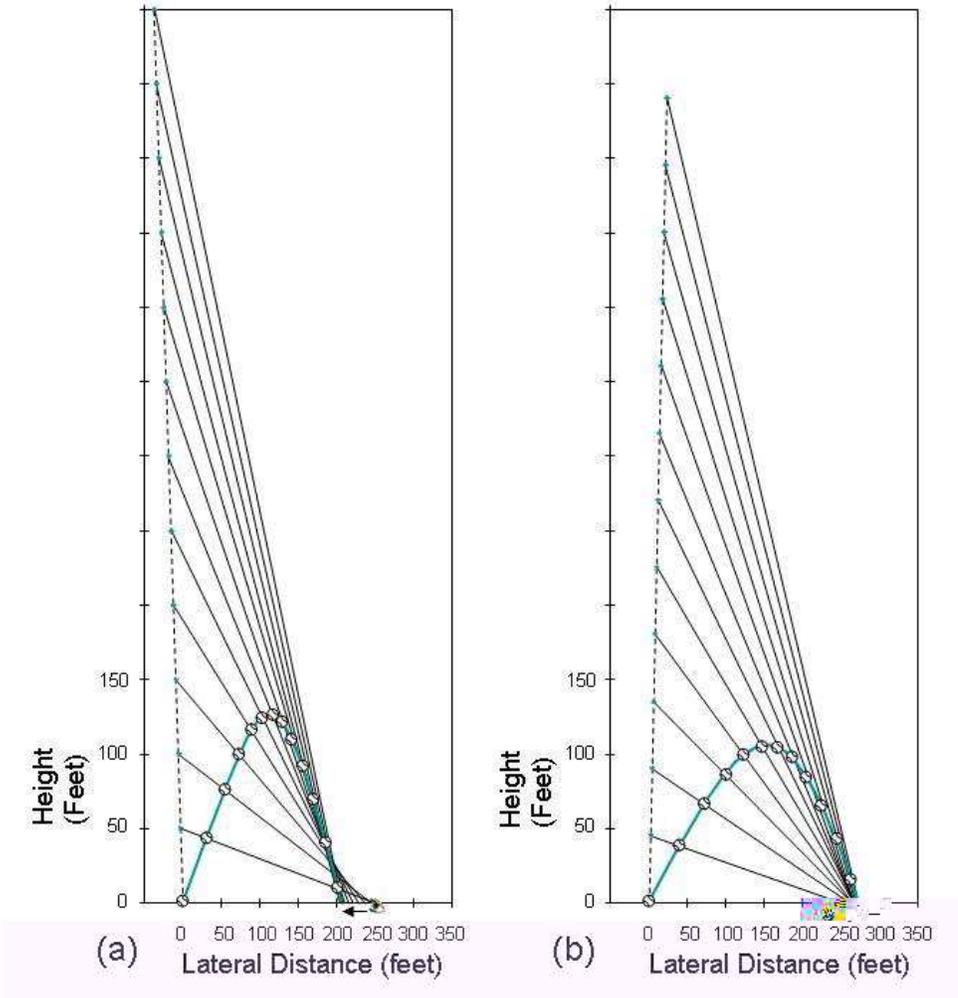} \caption{\small
Optical Acceleration Cancellation (OAC) with moving fielders and
realistically modelled trajectories as the interception control
mechanism for moving fielders. The outfielder starts at a distance
of 250 feet from home plate. OAC directs fielder to approach
desired destination along a smooth, monotonic running path. Shown
are side views of the ball moving from left to right, and fielder
moving from the picture of eyeball at the right. The left diagram
is the case of OAC directing fielder forward to catch a condition
$D=1.25$ trajectory. The right diagram is the case of OAC
directing fielder backward to catch a condition $D=1.0$
trajectory. In both cases, OAC produces a near constant running
velocity along the path to the ball.} \label{fig:oac}
\end{figure}

An example of how a fielder utilizes the OAC control strategy to
intercept a routine fly ball to the outfield is given in
Fig.~\ref{fig:oac}.  This figure illustrates the side view of a
moving fielder using OAC control strategy to intercept two
realistic outfield trajectories determined by our aerodynamics
model described in Sec.~\ref{sec:trajs}.  As specified by OAC, the
fielder simply runs up or back as needed to keep the tangent of
the vertical optical angle to the ball increasing at a constant
rate. Since the trajectory deviates from a parabola, the fielder
compensates by altering running speed somewhat. Geometrically the
OAC solution can be described as the fielder keeping the image of
the ball rising at a constant rate along a vertical projection
plane that moves forward or backwards to remain equidistant to the
fielder. For fly balls of this length, the geometric solution is
roughly equivalent to the fielder moving in space to keep the
image of the ball aligned with an imaginary elevator that starts
at home plate and is tilted forward or backward by the amount
corresponding to the distance that the fielder runs. As can be
seen in the figure, these outfield trajectories are notably
asymmetric, principally due to air resistance shortening, yet OAC
still guides the fielder along a smooth, monotonic running path to
the desired destination. This simple, relatively direct
navigational behavior has been observed in virtually all previous
perception-action catching studies with humans and
animals.\cite{shafer04}

\subsection{Application to examples of paradoxical trajectories}
\label{sec:stategies}
 Most previous models of interceptive
perception-action assume that real-world fly ball trajectories
remain similar enough to parabolic for robust optical control
strategies like OAC to generally produce simple, monotonic running
path solutions. Supporting tests have confirmed simple behavior
consistent with OAC in relatively extreme interception conditions
including catching curving Frisbees, towering outfield blasts and
short infield
pop-ups.\cite{mcbeath95,mcleod96,sugar06a,shafer04,sugar06b} The
apparent robustness of these optical control mechanisms implies
the commonly observed vacillating and lurching of fielders
pursuing high pop-ups must be due to some inexplicable cause. It
appears that the infielder is an unfortunate victim of odd wind
conditions, if not perhaps a bit too much chew tobacco or a nip of
something the inning before. In the current work, we have provided
evidence that there is a class of high infield pop-ups that we
refer to as paradoxical. Next we show that these deviate from
normal parabolic shape in ways dramatic enough to lead fielders
using OAC to systematically head off in the wrong direction or bob
forward and back. Below we illustrate how a fielder guided by OAC
will behave with each of the three paradoxical pop fly
trajectories that we determined in Sec.~\ref{sec:simulations} of
this paper.

\begin{figure}[ht]
\epsfig{width=4in,angle=270,file=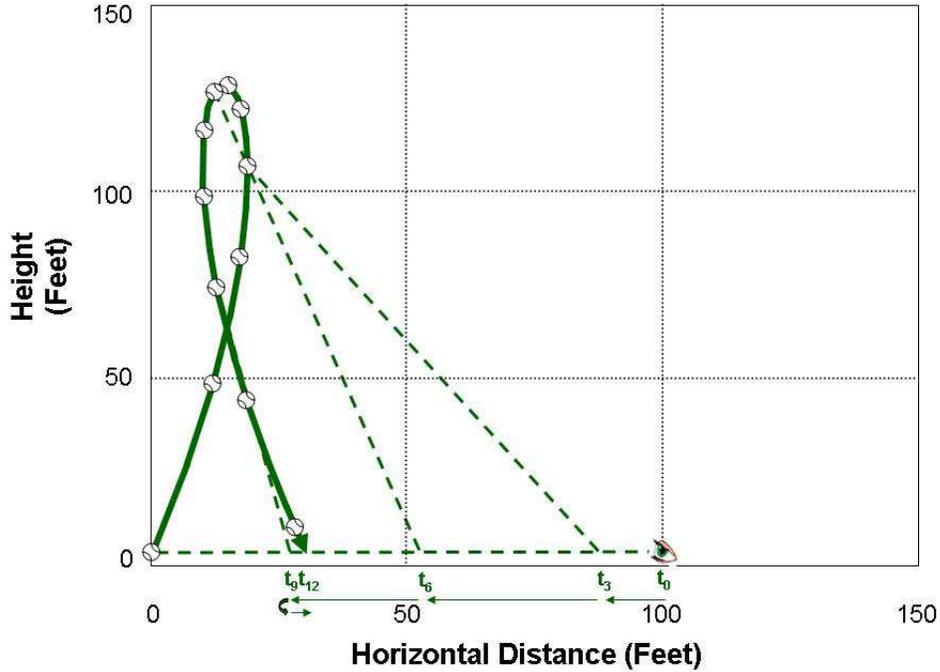} \caption{\small
Side view of the horizontal and vertical trajectory of a pop-up,
e. g. the trajectory seen by the first baseman watching a pop-up
being fielded by the third baseman. The dashed lines show the
fielder's gaze from his present position to the ball's present
position. The ``balls" show the trajectory at half-second
increments. The ``eye" shows the fielder's position at the start
of the trajectory. The fielder moves in the direction shown by the
arrows and exhibits little change in direction with a brief
potential back turn near the end.  This trajectory is for a
$D=1.7$ pop-up.} \label{fig:pop17}
\end{figure}

We first examine perhaps the most extreme paradoxical trajectory
of the group, the case of $D=1.7$, shown in Fig.~\ref{fig:pop17}.
This trajectory actually does a full loop-the-loop between the
catcher and pitcher, finally curving back out on its descent and
landing about 30 feet from home plate. Given the extreme
directional changes of this trajectory, we might expect an
infielder beginning 100 feet from home plate to experience
difficulty achieving graceful interception. Yet, as can be seen in
the figure, this case actually results in a relatively smooth
running path solution. When the fielder maintains OAC throughout
his approach, he initially runs quickly forward, then slightly
overshoots the destination, and finally lurches back. In practice,
near the interception point, the fielder is so close to the
approaching ball that it seems likely the eventual availability of
other depth cues like stereo disparity and rate of change in
optical size of the ball will mitigate any final lurch, and result
in a fairly smooth overall running path to the destination.

\begin{figure}[ht]
\epsfig{width=4in,angle=270,file=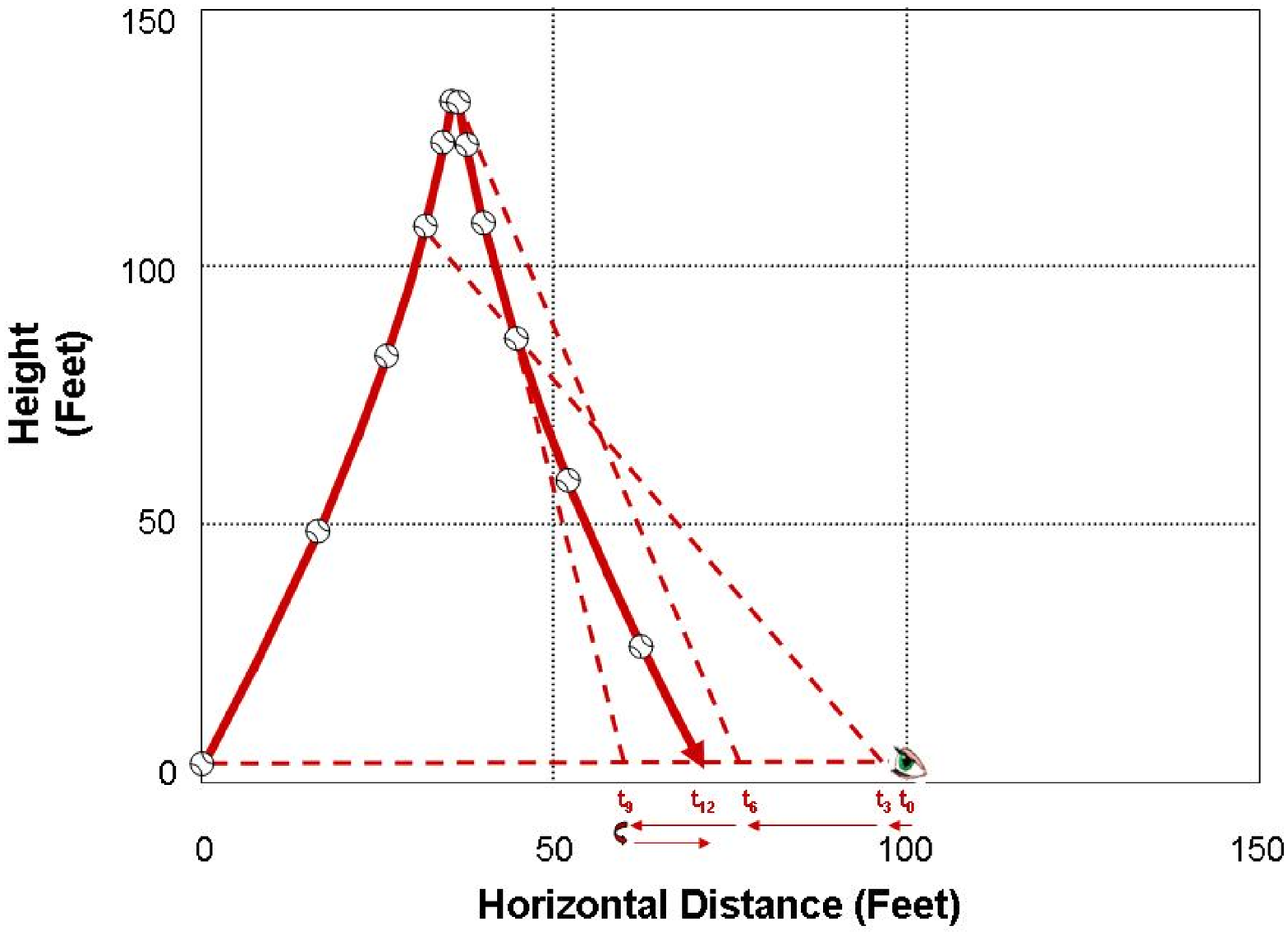} \caption{\small
Side view of fielder using OAC to run up to a pop fly from the
$D=1.6$ condition. Here the fielder dramatically changes direction
near the end.} \label{fig:pop16}
\end{figure}

Second we examine the case of a pop fly resulting from a bat-ball
offset $D=1.6$ in Fig.~\ref{fig:pop16}. Here the horizontal
velocity decreases in the beginning and approaches zero velocity
near the apex. Then after the apex, the Magnus force increases the
horizontal velocity. Yet, of greater impact to the fielder is that
this trajectory's destination is near where the fielder begins.
Thus from the fielder's perspective, before the discontinuity
takes place the trajectory slows in the depth direction such as to
guide the fielder to run up too far and then later to reverse
course and backtrack to where the ball is now accelerating
forward. Here the normally reliable OAC strategy leads the fielder
to systematically run up too far and in the final second lurch
backwards.

\begin{figure}[ht]
\epsfig{width=4in,angle=270,file=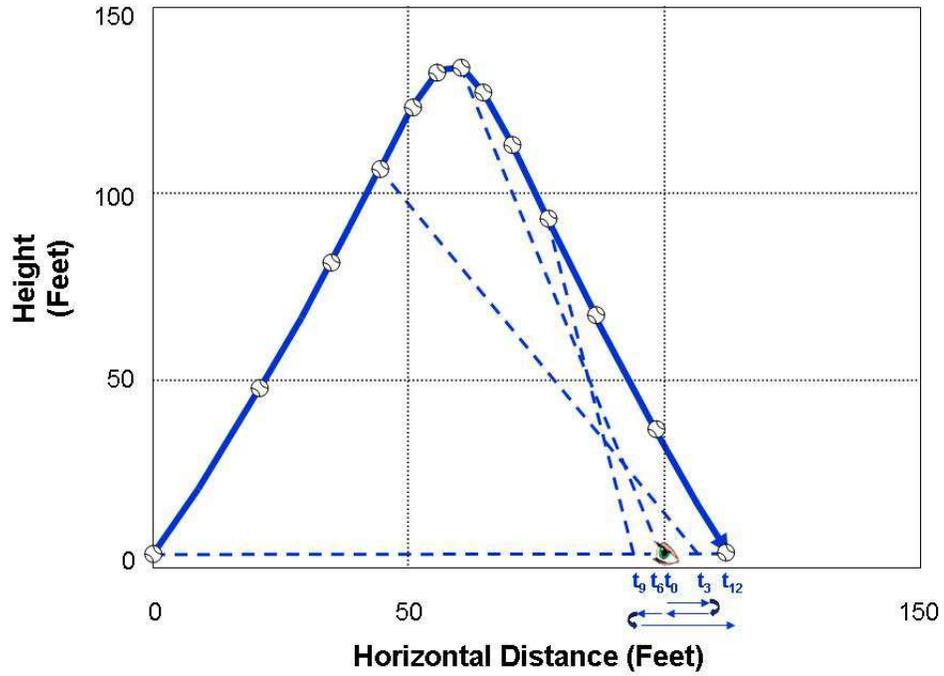} \caption{\small
Side view of fielder using OAC to run up to a pop fly from the
$D=1.5$ condition.  Here the fielder actually changes direction
twice, initially heading back, then forward, and finally back
again.} \label{fig:pop15}
\end{figure}

Third, we examine the case of a pop fly that lands just beyond the
fielder, the $D=1.5$ condition, in Fig.~\ref{fig:pop15}. In this
case OAC leads the fielder to initially head back to very near
where the ball is headed, but then soon after change direction and
run forward, only to have to run back again at the end. Certainly,
when a fielder vacillates or ``dances around" this much, it does
not appear that he is being guided well to the ball destination.
Yet, this seemingly misguided movement is precisely specified by
the OAC control mechanism. Thus, the assumption that fielders use
OAC leads to the bold prediction that even experienced,
professional infielders are likely to vacillate and make a final
lurch backward when navigating to catch some high, hard-hit
pop-ups, and indeed this is a commonly witnessed phenomenon.
Former major league infielders have affirmed to us that pop-ups
landing at the edge of the outfield grass (100 to 130 ft. from
home plate) usually are the most difficult to catch.

It is notable that in each of the cases depicted in
Figs.~\ref{fig:pop17}-\ref{fig:pop15}, the final movement by the
fielder prior to catching the ball is backwards.  This feature can
be directly attributed to the curvature of the trajectory, as
discussed in Sec.~\ref{sec:trajs}.  For a typical fly ball, the
curvature is small and negative, so the ball breaks slightly
towards home plate as it nears the end of its trajectory.  For
pop-ups, the curvature is large and positive, so the ball breaks
away from home plate, forcing the fielder to move backward just
prior to catching the ball.

\section{Summary and conclusions}
\label{sec:concl}

Why are very high pop-ups so hard to catch?  Using models of the
bat-ball collision and ball flight aerodynamics, we have shown
that the trajectories of these pop-ups have unexpected features,
such as loops and cusps.  We then examined the running paths that
occur with these dramatically non-parabolic trajectories when a
fielder utilizes OAC, a control strategy that has been shown
effective for tracking near-parabolic trajectories.  The predicted
behavior is very similar to observed behavior of infielders
attempting to catch high pop-ups.  They often vacillate forward
and backward in trying to position themselves properly to make the
catch, and frequently these changes in direction can lead to
confusion and positioning error.  Former major league infielders
confirm that our model agrees with their experiences.

\section*{Acknowledgments}
We are grateful to former major league players Clete Boyer, Jim
French, Norm Gigon, Bill Heath, Dave Hirtz, and Wayne Terwilliger
for their valuable comments and advice. Also, we thank David W.
Smith and Stephen D. Boren for information they provided about
pop-ups in the major leagues. Finally, we thank Bob Adair for
sharing his own unpublished work with us on judging fly balls and
for the insight regarding the final backward movement.

\pagebreak

\newpage

\vs
\vs

\newpage

%\begin{figure}[ht]
%\caption{\small Chapman's illustration of Optical Acceleration Cancellation (OAC) control mechanism.}
%\label{fig:chapman}
%\end{figure}

%\begin{figure}[ht]
%\caption{\small Linear Optical Control Mechanism (LOT) used to guide fielders to balls headed to the side.}
%\label{fig:lot}
%\end{figure}

\end{document}